\documentstyle[12pt,epsfig]{article}

\begin{document}
\baselineskip=23pt

\vspace{1.2cm}

\begin{center}
{\huge \bf  New Geometric Formalism for Gravity Equation in Empty
Space}

\bigskip

Xin-Bing Huang\footnote{huangxb@pku.edu.cn}\\
{\em Department of Physics,
Peking University,} \\
{\em  100871 Beijing, P. R. China}
\end{center}

\bigskip
\bigskip
\bigskip

\centerline{\large Abstract} In this paper, a complex daor field
which can be regarded as the square root of space-time metric is
proposed to represent gravity. The locally complexified geometry
is set up, and the complex spin connection constructs a bridge
between gravity and SU(1,3) gauge field. Daor field equations in
empty space are acquired, which are one-order differential
equations and not conflict with Einstein's gravity theory.

\vspace{1.2cm}


\vspace{1.2cm}

\newpage

\section{ Introduction\label{sec:Sec1}}

How to incorporate general relativity with quantum field theory is
a big problem in modern physics. There are now several popular
paradigms trying to solve this problem, among which string theory
and loop quantum gravity are well-known. The string theory
inherits most merits of quantum field theory. The graviton in
string theory is a spin-2 bosonic string. Since string theory is
constructed in 10-dimensional space-time (11-dimensional
space-time for M-theory), there are enough inner freedom to
accommodate all gauge interactions~\cite{pol98}. The most
difficult question in string theory is how to compactify
extra-dimensions to predict new observable phenomena. The loop
quantum gravity theory originates from Ashtekar's re-expression
for general relativity~\cite{ash86}, and inherits the geometric
viewpoint from general relativity. The loop quantum gravity theory
has yielded several interesting results, such as the discrete
spectrum of the space volume~\cite{tal01}.

Based on the concept of Yang-Mills field~\cite{yan54} and the
spontaneous symmetry broken mechanism~\cite{hig64}, Glashow, Salam
and Weinberg constructed a renormalizable electroweak gauge
theory~\cite{gsw60}. Combining this theory with quantum
chromodynamics yielded the so-called $SU(3)_{C}\times
SU(2)_{L}\times U(1)_{Y}$ standard modern in particle physics.
Recently, it is claimed that some great unified theories (GUTs)
endowed with SO(10) gauge group~\cite{bm93} can predict neutrino
masses and mixing angles~\cite{xin04} which are not contradictory
to modern experiments~\cite{yan04}. If physicists believe that the
standard model or GUTs should be a low-energy approximation of a
high energy unified quantum theory which incorporate gravity with
gauge fields, then those inner symmetry such as $SU(3)_{C}$ and
$U(1)_{Y}$ should be reasonably accommodated in the higher energy
theory. In this paper we propose a possibility to give those inner
symmetries without introducing extra-dimensions. Then in our
framework there are no difficulties on compactification. After
proposing the basic principles we give the locally complexified
geometry and study the doar field equations in a simple
case$--$gravity in empty space. Then we construct new geometric
formalism for gravity equation in empty space.

In general relativity Einstein made the assumption that gravity
equation in empty \footnote{Here ``empty" means that there is no
matter present and no physical fields except the gravitational
field. } space is~\cite{mtw73,dir75}
\begin{equation}
\label{daor1} R_{\mu\nu}=0~,
\end{equation}
where $R_{\mu\nu}$ is Ricci tensor, which is symmetrical.

In Minkowski space-time, Dirac equation is usually written
as($\hbar=c=1$)~\cite{dir58}
\begin{equation}
\label{daor2} \left(i\gamma^{a}\frac{\partial}{\partial
x^{a}}-m\right)\psi(x)=0~.~~~~~~a=0,1,2,3.
 \end{equation}
Where $\gamma$'s are Dirac matrices, which satisfy
\begin{equation}
\label{daor201}
\gamma^{a}\gamma^{b}+\gamma^{b}\gamma^{a}=2\eta^{ab}~.
\end{equation}
Where $\eta^{ab}$ is the metric tensor of Minkowski space-time,
i.e.
\begin{eqnarray}\label{daor202}
\eta^{00}=+1~,~~~\eta^{11}=\eta^{22}=\eta^{33}=-1~,~~~
 \eta^{ab}=0~~~~{\rm for}~~~~a \neq b~.
\end{eqnarray}
Eq.(\ref{daor2}) describes a free massive spinor field.

Physicists will find many advantages in Eq.(\ref{daor2}) when he
tries to realize the quantization of fields. Dirac equation is a
one-order differential equation. The spinor $\psi(x)$ is complex,
which is consistent with the wave function in Schr$\ddot{\rm
o}$dinger equation in quantum mechanics. Comparing Einstein
equation (\ref{daor1}) with Dirac equation (\ref{daor2}), we
believe that quantizing gravitation needs to reformulate gravity
theory such that new formalism at least includes two properties:
(1)the complexified field; (2)the reduction of the order of the
field equation.

In this paper we propose a basic principle that inner symmetries
arise from the extended local symmetry of space-time. The
complexified vierbeins (or tetrads) should be treated as the
fundamental field. Then we give the locally complexified geometry
and find that SU(1,3) Yang-Mills field appears as a natural result
without introducing extra-dimension.  Furthermore we give a new
geometric formalism for gravity equation in empty space which
looks more like Dirac equation. At the end we conclude the idea on
daor field and the main results of this paper.

\section{Basic Principles\label{sec:Sec2}}

\renewcommand{\theequation}{2.\arabic{equation}}
\setcounter{equation}{0}

At first we propose two basic principles: (1)Our space-time is a
$3+1$ manifold, which looks like a Minkowski space-time around
each point. (2)The intrinsic
distance\footnote{$x^{\mu}$'s($\mu=0,1,2,3$) are a system of
curvilinear coordinates of space-time manifold. }
\begin{equation}
\label{daor3} ds^2=dx^{\mu}g_{\mu\nu}dx^{\nu}
 \end{equation}
is invariant under any physical transformations. One kind of these
local transformations is corresponding to one kind of
interactions.

In the literature, decomposing the curvilinear metric into
vierbeins or tetrads $e^{a}_{~\mu}(x)$ has been used
extensively~\cite{dir58,uti56}. But the vierbein
decomposition\footnote{In this paper, using Roman suffixes to
refer to the bases of local Minkowski frame; using Greek suffixes
to refer to the space-time coordinates.}
\begin{equation}
\label{daor301} g_{\mu\nu}=\eta_{ab}e^{a}_{~\mu}(x)e^{b}_{~\nu}(x)
 \end{equation}
just keeps $ds^{2}$ invariant under local SO$(1,3)$ group
transformation, which is not the largest symmetry group
transformation as we will show. Now we will discuss the larger
symmetry transformation which keeps $ds^{2}$ invariant. To do so,
we introduce a complex vierbein field $h^{a}_{~\mu}$ or
$H^{~\mu}_{a}$, which satisfies
\begin{eqnarray}
\label{daor401}
2g_{\mu\nu}&=&\bar{h}^{~a}_{\mu}\eta_{ab}h^{b}_{~\nu}
+h^{~a}_{\mu}\eta_{ab}\bar{h}^{b}_{~\nu}~,
\\
\label{daor402}
2g^{\mu\nu}&=&{\bar{H}}^{\mu}_{~a}\eta^{ab}H^{~\nu}_{b}
+H^{\mu}_{~a}\eta^{ab}{\bar{H}}^{~\nu}_{b}~,
\\
\label{daor403}
g_{\mu\nu}g^{\nu\lambda}&=&g^{\lambda\nu}g_{\nu\mu}=\delta^{\lambda}_{\mu}~,
\end{eqnarray}
where bar denotes complex conjugation. It is stressed that
$g_{\mu\nu}$ is still real and symmetrical in above equations.
About half a century ago, Einstein and Strauss constructed their
theory under a complexified geometry where the metric is hermitian
and thus asymmetry~\cite{ein46}. After them Schr$\ddot{\rm
o}$dinger explained this theory explicitly in his famous
book~\cite{sch85}. Following the method of complexified geometry,
Penrose proposed the method of twistor to argue the quantization
of space-time~\cite{pen62}. The main development on this way
before 1980s was reviewed by Israel~\cite{isr79}. In the process
of preparing this paper, we found that Ali H. Chamseddine had used
complex vierbein to construct gravity theory under noncommutative
geometry~\cite{cha04}.

At the first blush, Eq.(\ref{daor401}) looks like the following
formula\footnote{Since Fock first used this relation to study the
Dirac equation on Riemannian space-time~\cite{foc27}, this
equation had been always adopted by many physicists.}
\begin{eqnarray}
\label{fock} 2g_{\mu\nu}(x)&=&\gamma_{\mu}(x)\gamma_{\nu}(x)
+\gamma_{\nu}(x)\gamma_{\mu}(x)~.
\end{eqnarray}
Similar to Eq.(\ref{daor201}) it is obvious that $\gamma_{\mu}(x)$
in Eq.(\ref{fock}) should be matrices in the spin space. But
$h^{b}_{~\nu}(x)$ in Eq.(\ref{daor401}) don't relate to the spin
space. The complex vierbein and complex spin connection have
appeared in twistor theory~\cite{pen62} or in some formulae in
loop quantum gravity~\cite{tal01}. We directly use the
complexification of vierbein to reformulate the gravity equation
in empty space and argue the possibility of accommodating SU(1,3)
Yang-Mills field in our framework in this paper.

To embody the linking character of $h^{a}_{~\mu}(x)$ between
matter fields and the space-time structure, also between
gravitation and gauge interactions, here we intend to give the
complex vierbein (or tetrad) field $h^{a}_{~\mu}(x)$ a new name.
``Dao" is a basic and important concept in ancient Chinese
philosophy. ``Dao" is used to refer not only the unobservable
existence from which everything originate but also the laws which
dominate the doom of everything. ``Dao" is also used to
demonstrate the abstract relationship between the dual things such
as ``Yin" and ``Yang", nihility and existence. Since
$h^{a}_{~\mu}(x)$ plays such a similar role in physics, which will
be discussed in the following, we suggest calling
$h^{a}_{~\mu}(x)$ the ``daor field".

The metric is symmetrical in general relativity. Thus, there are
at most 10 free-parameters in curvilinear metric $g_{\mu\nu}$.
Eq.(\ref{daor401}) and Eq.(\ref{daor402}) demonstrate that there
are too much free-parameters in the daor field. To cancel
nonphysical freedom, we require that the following covariant
constraint should be satisfied
\begin{equation}
\label{daor-con}
\bar{h}^{~a}_{\mu}\eta_{ab}h^{b}_{~\nu}=h^{~a}_{\mu}\eta_{ab}\bar{h}^{b}_{~\nu}
~.
\end{equation}
It is obvious that Eq.(\ref{daor-con}) can be rewritten
equivalently
\begin{equation}
\label{daor-con++} \bar{H}^{\mu}_{~a}\eta^{ab}H^{~\nu}_{b}
=H^{\mu}_{~a}\eta^{ab}\bar{H}^{~\nu}_{b} ~.
\end{equation}
Then Eq.(\ref{daor401}) and Eq.(\ref{daor402}) become
\begin{equation}
\label{daor4} g_{\mu\nu}=\bar{h}^{~a}_{\mu}\eta_{ab}h^{b}_{~\nu}~
,~~~~~~g^{\mu\nu}=\bar{H}^{\mu}_{~a}\eta^{ab}H^{~\nu}_{b}~.
\end{equation}

Define a tensor $N_{\mu\nu}$ to be
\begin{equation}
\label{zero-matrix}
2N_{\mu\nu}=\bar{h}^{~a}_{\mu}\eta_{ab}h^{b}_{~\nu}
-h^{~a}_{\mu}\eta_{ab}\bar{h}^{b}_{~\nu}~.
\end{equation}
Eq.(\ref{zero-matrix}) shows that the tensor $N_{\mu\nu}$ is
antisymmetrical, namely $N_{\mu\nu}=-N_{\nu\mu}$. Thus the
covariant constraint $N_{\mu\nu}=0$ only provides 6 independent
constraint equations to the components of the daor field. That is
to say, the daor field can at most have 26 independent
free-parameters.

Consider general real coordinate transformations $x\rightarrow
x^{\prime}(x)$, since
\begin{equation}
\label{co-trans} dx^{\prime\mu}=\frac{\partial x^{\prime\mu}
}{\partial x^{\nu}}dx^{\nu}~ ,~~~~~~h^{\prime
b}_{~~\mu}=h^{b}_{~\nu}\frac{\partial x^{\nu}}{\partial
x^{\prime\mu} }~,
\end{equation}
the intrinsic distance is thus invariant under general coordinate
transformations.

Under the rotation of the locally complexified Minkowski frame,
the daor field $h^{b}_{~\nu}(x)$ transforms as follows
\begin{equation}
\label{ro-trans00} h^{a}_{~\nu}(x)\rightarrow h^{\prime
a}_{~\nu}(x)=S^{a}_{~b}(x)h^{b}_{~\nu}(x)~.
\end{equation}
if the matrix $S^{a}_{~b}(x)$ satisfies
\begin{equation}
\label{ro-trans01}
\bar{S}^{~c}_{a}(x)\eta_{cd}S^{d}_{~b}(x)=\eta_{ab}~,
\end{equation}
namely, $S^{a}_{~b}(x)$ being the element of U(1,3) group, then
the intrinsic distance is invariant under the rotation of the
local complexified Minkowski frame.

Hence we can draw a conclusion: by introducing the complex daor
field, we find that the intrinsic distance and the covariant
constraint are invariant under two kinds of transformations: One
is the general coordinate transformation $x\rightarrow
x^{\prime}(x)$; The other is the U(1,3) transformation of the
Roman suffixes.


\section{Daor Geometry\label{sec:Sec3}}

\renewcommand{\theequation}{3.\arabic{equation}}
\setcounter{equation}{0}

It is well known that the aim of geometry is to study the
invariant properties of the continuous point set $E$ (called
space) under some symmetry transformations. So, from the geometric
point of view, there are also a kind of geometry in which the
distance defined by Eq.(\ref{daor3}) is invariant under two kinds
of symmetry transformations discussed in the last section.

By defining the Hermitian conjugate of the daor field
$h^{a}_{~\mu}$ or $H^{~\mu}_{a}$
\begin{equation}
\label{daor5} h^{\dag}=(h^{a}_{~\mu})^{\dag}=(\bar{h}^{~a}_{\mu})
~,~~H^{\dag}=(H^{~\mu}_{a})^{\dag}=(\bar{H}^{\mu}_{~a})~,
\end{equation}
from Eq.(\ref{daor403}) and Eq.(\ref{daor4}) we can easily acquire
the following relations
\begin{eqnarray}\label{daor6}
 g=h^{\dag}\eta h~ ,~~~~~~~~g^{-1}=H^{\dag}\eta H~,~~~~~~~~
 H^{\dag}&=&h^{-1}~,~~~~~~~~H^{-1}=h^{\dag}~.
\end{eqnarray}

To provide an algebraic preparation for an advanced study in daor
manifold, we will study multilinear complex algebra without
considering the covariant constraint first. Now let's define
\footnote{We use $\partial_{\mu}$ to denote the partial
differential operator $\frac{\partial}{\partial x^{\mu}}$ for
brevity.}
\begin{eqnarray}\label{daor4001}
 h^{a}=h^{a}_{~\mu}{\rm d}x^{\mu}~,~~~~~h^{\dag
a}={\rm d}x^{\mu} \bar{h}^{~a}_{\mu}~,~~~~~~
{H}^{\dag}_{a}=\bar{H}^{\mu}_{~a}\partial_{\mu}~,
~~~~~~H_{a}=H^{~\mu}_{a}\partial_{\mu}~,
\end{eqnarray}
where $h^{a}$ and $h^{\dag a}$ are daor field 1-forms. Here we
choose the set $\{
{H}^{\dag}_{0},{H}^{\dag}_{1},{H}^{\dag}_{2},{H}^{\dag}_{3} \}$ to
be the basis of locally complexified Minkowski frame because they
are linearly independent everywhere. Thus any vector of locally
complexified Minkowski space-time ${\bf M}$ can be uniquely
expressed by a linear combination of ${H}^{\dag}_{a}$'s. Suppose
$f:{\bf M}\longrightarrow {\mathbf C}$ is a ${\mathbf
C}$-valued\footnote{${\mathbf C}$ refers to the field of complex
numbers in this paper.} function on ${\bf M}$. If for any $Z_{1},
Z_{2}\in {\bf M}$ and $\alpha^{1},\alpha^{2} \in {\mathbf C}$,
\begin{equation}
\label{daor4002} f(\alpha^{1}Z_{1}+\alpha^{2}Z_{2})=\alpha^{1}
f(Z_{1})+\alpha^{2} f( Z_{2})~,
\end{equation}
then $f$ is called a ${\mathbf C}$-valued linear function on ${\bf
M}$. It is easy to prove that the set of all ${\mathbf C}$-valued
linear functions on ${\bf M}$ forms a vector space over ${\mathbf
C}$, called the dual space of ${\bf M}$~\cite{ccl99}, denoted by
${\bf M}^{*}$. The Hermitian conjugated local Minkowski space-time
is denoted by $ {\bf M}^{\dag}$, in which the set $\{
H_{0},H_{1},H_{2},H_{3} \}$ is a basis.

By using the same definition of inner product as in real manifold,
namely
\begin{equation}
\label{daor4003} <\partial_{\mu},{\rm
d}x^{\nu}>=\delta^{\nu}_{\mu}~,
\end{equation}
from Eq.(\ref{daor6}) and Eq.(\ref{daor4003}), we can obtain the
inner product of $h^{a}$ and ${H}^{\dag}_{a}$ as follows
\begin{equation}
\label{daor6001} <{H}^{\dag}_{a},h^{b}>=<H_{a},{h}^{\dag
b}>=\delta^{b}_{a}~.
\end{equation}
Eq.(\ref{daor6001}) demonstrates that the set $\{
h^{0},h^{1},h^{2},h^{3} \}$ is a basis of ${\bf M}^{*}$.
Obviously, ${\bf M}^{*}$ and ${\bf M}$ are dual spaces of each
other.

Furthermore, the inner product of the vector $U=U^{\mu}
{\partial}_{\mu}$ and the covector $V=V_{\nu}{\rm d}x^{\nu}$ can
be expressed as follows
\begin{equation}
\label{daor604}
<U,V>=<V,U>=U^{\mu}V_{\mu}=V_{\mu}U^{\mu}=V_{a}U^{a}={\bar
U^{a}}{\bar V}_{a}~.
\end{equation}
Where ${\bar U}^{a}$, $U^{a}$, ${\bar V}_{a}$ and $V_{a}$ are
given by
\begin{eqnarray}\label{daor605}
 U^{a}=h_{~\mu}^{a}U^{\mu}~,~~~~{\bar
U}^{a}=U^{\mu}\bar{h}_{\mu}^{~a}~,~~~~ V_{a}=V_{\mu}{\bar
H}_{~a}^{\mu}~,~~~~{\bar V}_{a}=H_{a}^{~\mu}V_{\mu}~.
\end{eqnarray}
For brevity, we introduce the signs $U^{\bar {a}}$ and $V_{\bar
{a}}$ defined by
\begin{equation}
\label{daor606} U^{\bar {a}}={\bar
U}^{a}=U^{\mu}\bar{h}_{\mu}^{~a}~,~~~~~~V_{\bar {a}}={\bar
V}_{a}=H_{a}^{~\mu}V_{\mu}~.
\end{equation}
The vector $U=U^{a}{H}^{\dag}_{a}$ and the covector $V= V_{a}
h^{a}$  are obviously invariant under two kinds of
transformations.

Next let's give the notion of tensors. The elements in the tensor
product
\begin{equation}
\label{tensor001} {\bf M}^{r}_{s}=\overbrace{{\bf M} \otimes
\cdots \otimes {\bf M}}^{r~ terms} \otimes \overbrace{{\bf M}^{*}
\otimes \cdots \otimes {\bf M}^{*}}^{s~ terms}~,
\end{equation}
are called $(r,s)$-type tensors, where $r$ is the contravariant
order and $s$ is the covariant order. The elements in ${\bf
M}^{r}_{0}$ are called contravariant tensors of order $r$, and
those in ${\bf M}^{0}_{s}$ are called covariant tensors of order
$s$. We also use the following conventions: ${\bf
M}^{0}_{0}={\mathbf C}$, ${\bf M}^{1}_{0}={\bf M}$, ${\bf
M}^{0}_{1}={\bf M}^{*}$. Since $\{ {H}^{\dag}_{i} \}_{0\leq i\leq
3}$ and $\{ {h}^{i} \}_{0\leq i\leq 3}$ are dual bases in ${\bf
M}$ and ${\bf M}^{*}$ respectively, then
\begin{eqnarray}\label{tensor002}
 {H}^{\dag}_{i_1} \otimes \cdots \otimes
{H}^{\dag}_{i_r} \otimes h^{j_{1}} \otimes \cdots \otimes
h^{j_{s}}~,~~~~
 0\leq i_1,\ldots,i_r,j_1,\ldots,j_s\leq 3~,
\end{eqnarray}
form a basis of ${\bf M}^{r}_{s}$. Therefore an $(r,s)$-type
tensor $T$ can be uniquely expressed as
\begin{equation}
\label{tensor003} T=T^{{i}_1\cdots {i}_r}_{j_1 \cdots j_s}
{H}^{\dag}_{i_1} \otimes \cdots \otimes {H}^{\dag}_{i_r} \otimes
h^{j_{1}} \otimes \cdots \otimes h^{j_{s}}~,
\end{equation}
where the $T^{{i_1}\cdots {i_r}}_{j_1 \cdots j_s}$ are called the
components of the tensor $T$ under the basis (\ref{tensor002}).

In the curvilinear coordinate a real $(r,s)$-type tensor $P$ is
uniquely expressed as
\begin{equation}
\label{tensor004} P=P^{\mu_1\cdots \mu_r}_{\nu_1 \cdots \nu_s}
 \partial_{\mu_1} \otimes \cdots \otimes \partial_{\mu_r} \otimes
{\rm d}x^{\nu_{1}} \otimes \cdots \otimes {\rm d}x^{\nu_{s}}~.
\end{equation}
From Eq.(\ref{daor4001}) and Eq.(\ref{daor605})  we therefore
acquire
\begin{equation}
\label{tensor005} P^{\mu_1\cdots \mu_r}_{\nu_1 \cdots \nu_s}=P^{
{i}_1\cdots {i}_r }_{j_1 \cdots j_s} {\bar H}^{\mu_1}_{~i_1}
\cdots {\bar H}^{\mu_r}_{~i_r} h_{~\nu_1}^{j_1}\cdots
h_{~\nu_s}^{j_s}~.
\end{equation}
Suppose $J$ is an $(r_1,s_1)$-type tensor and $K$ is an
$(r_2,s_2)$-type tensor. Then their tensor product $J \otimes K$
is an $(r_1+r_2,s_1+s_2)$-type tensor, and the components of $J
\otimes K$ are the products of the components of $J$ and $K$,
i.e.,
\begin{equation}
\label{tensor006} (J \otimes K)^{{i}_1\cdots {i}_{r_1+r_2} }_{j_1
\cdots j_{s_1+s_2} }=J^{{i}_1\cdots {i}_{r_1} }_{j_1 \cdots
j_{s_1} } K^{{i}_{r_1+1}\cdots {i}_{r_1+r_2} }_{j_{s_1+1} \cdots
j_{s_1+s_2} }~.
\end{equation}
The multiplication of tensors satisfies the distributive and
associative laws.

A special class of tensors, the totally skew-symmetric covariant
tensors have played an important role in the study of manifolds.
We begin by defining Cartan's exterior product as the
antisymmetric tensor product of cotangent space basis elements
$h^{a}$. For instance, the exterior product of covector fields,
which is a skew-symmetric linear mapping, called 2-form and
constitute a space $\Lambda^{2}({\bf M})$. The bases of the 2-form
space $\Lambda^{2}({\bf M})$ are
\begin{equation}
\label{wedge001}  h^{a}\wedge h^{b}=(h^{a} \otimes h^{b}-h^{b}
\otimes h^{a})~.
\end{equation}
Then any 2-form $\alpha_2\in \Lambda^{2}({\bf M})$ can be written
as
\begin{equation}
\label{wedge002}  \alpha_2=\frac{1}{2}f_{ab}h^{a}\wedge
h^{b}~,~~~~~~f_{ab}=-f_{ba}~.
\end{equation}
Let $\Lambda^{r}({\bf M})~(r=0,\ldots,4)$ be the set of
skew-symmetric ${\bf M}^{0}_{r}$ tensors. This is a vector space
of dimension $\frac{4!}{r!(4-r)!}$. Therefore, any $r$-form
$\alpha_r \in \Lambda^{r}({\bf M})$ can be expressed as
\begin{equation}
\label{wedge003}  \alpha_r=\frac{1}{r!}f_{i_1 \cdots
i_r}h^{i_1}\wedge \cdots \wedge h^{i_r}~,
\end{equation}
where function $f_{i_1 \cdots i_r}$ is completely skew-symmetric
with respect to its subscripts, and $\Lambda^{1}({\bf M})={\bf
M}^{*}$ is the set of complex cotangent vector fields. Denote the
formal sum
$$\sum^{4}_{r=0} \Lambda^{r}({\bf M})$$ by $ \Lambda^{*}({\bf M})$.
Then $\Lambda^{*}({\bf M})$ is a $2^4$-dimensional vector space.
Let
\begin{equation}
\label{wedge004}  \alpha=\sum^{4}_{r=0}
\alpha_{r}~,~~~~~~\beta=\sum^{4}_{s=0} \beta_{s}~,
\end{equation}
where $\alpha_r \in \Lambda^{r}({\bf M})$, $\beta_s \in
\Lambda^{s}({\bf M})$. Define the exterior product of $\alpha$ and
$\beta$ by
\begin{equation}
\label{wedge005}  \alpha\wedge \beta=\sum^{4}_{r,s=0} \alpha_{r}
\wedge \beta_{s}~.
\end{equation}
Then $ \Lambda^{*}({\bf M})$ becomes an algebra with respect to
the exterior product, and is called the exterior algebra or
Grassman algebra of ${\bf M}$. Obviously, $\alpha_r$ and $\beta_s$
satisfy $\alpha_r\wedge\beta_s=(-1)^{rs}\beta_s\wedge \alpha_r$.

Using the tool of exterior product, we then rewrite the covariant
constraint as
\begin{equation}
\label{daor607} \eta_{ab}{\bar {h}}^{a}\wedge
h^{b}=0~,~~~~~~\eta_{ab}{\bar {H}}^{a}\wedge H^{b}=0~.
\end{equation}

Another useful tool for manipulating differential forms is the
exterior differentiation ${\rm d}$, which is a differential
operator within Cartan exterior algebra $ \Lambda^{*}({\bf M})$,
and defined by:
\begin{equation}
\label{diff001} {\rm d}:~\Lambda^{r}({\bf M})\longrightarrow
\Lambda^{r+1}({\bf M})~.
\end{equation}
For a complex function $f(x^{\mu})$, its exterior differentiation
is expressed as
\begin{equation}
\label{diff002} {\rm d}f=f_{,a}h^{a}~~(f_{,a}\equiv \frac{\partial
f}{\partial h^{a}})~.
\end{equation}
For a $p$-form $\alpha_p=\frac{1}{p!}f_{a_1 \cdots
a_p}h^{a_1}\wedge \cdots \wedge h^{a_p}$, its exterior
differentiation is defined as follows
\begin{equation}
\label{diff003} {\rm d}\alpha_p=\frac{1}{p!}f_{a_1 \cdots
a_p,k}h^{k} \wedge h^{a_1}\wedge \cdots \wedge h^{a_p}~.
\end{equation}
Let $\alpha_p\in\Lambda^{p}({\bf M})$, $\beta_q\in\Lambda^{q}({\bf
M})$. Making use of Leibniz rule for differentiation of functions,
it can be proved that
\begin{equation}
\label{diff004} {\rm d}(\alpha_p\wedge\beta_q)={\rm
d}\alpha_p\wedge\beta_q+(-1)^{p}\alpha_p\wedge{\rm d}\beta_q~.
\end{equation}

The inner product Eq.(\ref{daor6001}) can be extended to the inner
product between a vector field $Y$ and a $k$-form $\alpha_k$ as
follows
\begin{equation}
\label{inn001} i_{Y}\alpha_{k} \equiv
<Y,\alpha_{k}>~\in~\Lambda^{k-1}({\bf M})~.
\end{equation}
Where the operator $i_{Y}$ acts on the differential forms only.
Express the vector field $Y$ as $Y=\zeta^{i}H^{\dag}_{i}$, it is
easy to obtain
\begin{eqnarray}\label{inn002}
i_{Y}\alpha_{k} = <Y,\alpha_{k}>
 =\frac{1}{(k-1)!}\zeta^{a_1}f_{a_1 \cdots
a_k}h^{a_2}\wedge \cdots \wedge h^{a_k}~.
\end{eqnarray}
having constructed fundamental algebraic tools, we will study the
geometric properties of the locally complexified manifold in the
next section.


\section{Daor Connection \label{sec:Sec4}}
\renewcommand{\theequation}{4.\arabic{equation}}
\setcounter{equation}{0}

In the following, we will consider a special subset of the daor
field $h^{a}_{~\mu}(x)$ in which an element $k^{a}_{~\mu}(x)$ is
expressed in terms of the real vierbeins as
\begin{equation}
\label{conn001} k^{a}_{~\mu}(x)=l^{a}_{~b}(x)e^{b}_{~\mu}(x)~,
\end{equation}
where the matrix $l^{a}_{~b}(x)$ satisfies Eq.(\ref{ro-trans01}),
namely, $l^{\dag}~\eta ~l=\eta$. It is obvious that the matrix
$k^{a}_{~\mu}(x)$ satisfies the covariant constraint
Eq.(\ref{daor-con}).

Being similar with Eq.(\ref{daor4001}), It can be defined that
\begin{eqnarray}\label{conn002}
e^{a}=e^{a}_{~\mu}{\rm
d}x^{\mu}~,~~~k^{a}=l^{a}_{~b}e^{b}~,~~~k^{\dag a}=e^{b}
\bar{l}^{~a}_{b}~,~~~{K}^{\dag}_{a}=\bar{K}^{\mu}_{~a}\partial_{\mu}~,
~~~K_{a}=K^{~\mu}_{a}\partial_{\mu}~,
\end{eqnarray}
where $\bar{K}^{\mu}_{~a}$ and $K^{~\mu}_{a}$ are given by
\begin{equation}
\label{conn003} {K}^{\dag}=k^{-1}~, ~~~~~~K^{-1}=k^{\dag}~.
\end{equation}

In order to ``differentiate" vector fields on the space-time
manifold, we need to introduce a structure called the connection
on a vector bundle. When the linear connection is given on the
cotangent bundle, there is a continuous linear mapping between
sections of the tensor bundle, namely:
\begin{equation}
\label{conn004} \nabla:~{\bf M}^{r}_{s}\longrightarrow {\bf
M}^{r}_{s+1}~,~~~~~~J\in{\bf M}^{r}_{s}\longrightarrow \nabla J\in
{\bf M}^{r}_{s+1}~.
\end{equation}
$\nabla J$ is called the covariant differentiation of tensor field
$J$~\cite{hh97}. The mapping $\nabla$ satisfies

1)Linearity:
\begin{equation}
\label{conn005} \nabla(aJ+bJ^{\prime})=a\nabla J+b\nabla
J^{\prime}~,~~~~~~a,b\in{\bf C}~.
\end{equation}

2)Leibniz rule:
\begin{eqnarray}
\label{conn00500} \nabla(J \otimes J^{\prime})&=&(\nabla J)\otimes
J^{\prime} +J\otimes (\nabla J^{\prime})~,\\ \label{conn006}
\nabla<Y,\alpha_r>&=&<\nabla Y,\alpha_r> +<Y,\nabla \alpha_r>~.
\end{eqnarray}

3)For function $f\in {\bf M}^{0}_{0}$,
\begin{equation}
\label{conn007} \nabla f={\rm d} f~.
\end{equation}

4)For cotangent vector field $\alpha_1=f_a (x) k^{a}(x)$, when the
daor field is chosen,
\begin{equation}
\label{conn008} \nabla \alpha_1={\rm d} f_a\otimes k^{a}+f_a\nabla
k^{a} ~.
\end{equation}
From Eq.(\ref{conn008}), $\nabla \alpha_1$ can be calculated if
covariant differentiation $\nabla k^{a}$ of a daor field is given.
$\nabla k^{a}$ denote the infinitesimal variance of the daor field
$k^a$ at the neighborhood of a point and can be expressed as
\begin{eqnarray}\label{conn00444}
 \nabla{k}^{a}=(\nabla
l^{a}_{~b})e^{b}+l^{a}_{~b}(\nabla e^{b})
=-l^{a}_{~c}B^{c}_{~b}e^{b}-l^{a}_{~b}\theta^{b}_{~c}e^{c}
=-l^{a}_{~b}\omega^{b}_{~c}e^{c}~,
\end{eqnarray}
where
\begin{equation}
\label{conn010}
\omega^{a}_{~b}=B^{a}_{~b}+\theta^{a}_{~b}=-<K^{\dag}_{b},\nabla
k^{a}
>=\omega^{a}_{~bi}k^{i} ~.
\end{equation}
Because $\omega^{a}_{~b}$ are complex matrix valued 1-forms, we
suggest calling $\omega^{a}_{~b}$ ``daor connection" 1-forms. From
Eq.(\ref{conn008}) and Eq.(\ref{conn00444}), we obtain
\begin{eqnarray}\label{conn011}
 \nabla \alpha_1={\rm d} f_a ~l^{a}_{~b}~e^{b}-f_a~
l^{a}_{~b}~\omega^{b}_{~c}~e^{c} =l^{a}_{~b}~({\rm d} f_a
~\delta^{b}_{c}-f_a~\omega^{b}_{~c})~ e^{c} ~.
\end{eqnarray}

Using Eq.(\ref{daor6001}) and Eq.(\ref{conn006}), daor connections
on the tangent bundle can be induced from daor connections on the
cotangent bundle. Since $\{ K^{\dag}_{a} \}$ are the dual bases of
$\{k^a   \}$, it is easy to prove that
\begin{equation}
\label{conn016} <\nabla K^{\dag}_{b},k^{a}
>=-<K^{\dag}_{b},\nabla k^{a}
>=\omega^{a}_{~b}~,
\end{equation}
or equivalently, $\nabla
K^{\dag}_{b}=(l^{-1})^{c}_{b}~\omega^{d}_{c}~(e^{-1})^{\mu}_{d}~\partial_{\mu}$.
Therefore, the covariant differentiation of a tangent field
$Y=\zeta^{a}(x)K^{\dag}_{a}(x)$ is
\begin{equation}
\label{conn018} \nabla Y
~=~(l^{-1})^{b}_{a}~(d\zeta^{a}~\delta^{c}_{b}+
\zeta^{a}~\omega^{c}_{b})~(e^{-1})^{\mu}_{c}~\partial_{\mu}~.
\end{equation}
Similarly, covariant differentiation of any $(r,s)$-type tensor
fields can be carried out yielding $(r,s+1)$-type tensor fields.

It is well known that the $\theta^{a}_{~b}$ defined in
Eq.(\ref{conn00444}) is the spin connection introduced first by
Cartan. From the viewpoints of Yang-Mills gauge field, the daor
connection $\omega^{a}_{~b}$ is the field strength of
SU(1,3)$\times$SO(1,3) gauge field. Or equivalently, in the
language of differential geometry, the $\omega^{a}_{~b}$ is the
connection on SU(1,3)$\times$SO(1,3) principal bundle. The
curvature of this principal bundle thus is expressed as
\begin{equation}
\label{daor1903} \Omega^{a}_{~b}={\rm d
}\omega^{a}_{~b}+\omega^{a}_{~c}\wedge\omega^{c}_{~b}~.
\end{equation}
Since Eq.(\ref{conn010}), the curvature $\Omega^{a}_{~b}$ can be
written as
\begin{eqnarray}
\label{daor190301}
 \Omega^{a}_{~b}=
R^{a}_{~b}+F^{a}_{~b}+\theta^{a}_{~c}\wedge
B^{c}_{~b}+B^{a}_{~c}\wedge\theta^{c}_{~b}~,
\end{eqnarray}
where $R^{a}_{~b}$, $F^{a}_{~b}$ are the curvature of SO(1,3)
principal bundle and the curvature of SU(1,3) principal bundle
respectively. The definitions of $R^{a}_{~b}$ and $F^{a}_{~b}$ are
\begin{eqnarray}
\label{daor190303} R^{a}_{~b}&=&{\rm d
}\theta^{a}_{~b}+\theta^{a}_{~c}\wedge \theta^{c}_{~b}~,\\
\label{daor1902++} F^{a}_{~b}&=&{\rm d }B^{a}_{~b} + B^{a}_{~c}
\wedge B^{c}_{~b}~.
\end{eqnarray}

Introducing SU(1,3) gauge field $\tilde{B}=\frac{1}{i\lambda}B$
and its field strength $\tilde{F}=\frac{1}{i\lambda}F$, we acquire
the well-known relation
\begin{equation}
\label{daor1902aa} \tilde{F}^{a}_{~b}={\rm d }\tilde{B}^{a}_{~b}
+i\lambda \tilde{B}^{a}_{~c} \wedge \tilde{B}^{c}_{~b}~,
\end{equation}
where $\lambda$ is the coupling constant of the SU(1,3) gauge
field. In quantum field theory the freedom of gauge field is too
much to describe the physical system. Theorists must introduce
some kind of gauge fixing condition, such as Coulomb gauge in QED
or Landau gauge in QCD, to give the observable physical results.
In our framework we propose the gauge fixing condition as follows
\begin{eqnarray}
\label{gauge-fixing} \theta^{a}_{~c}\wedge
\tilde{B}^{c}_{~b}+\tilde{B}^{a}_{~c}\wedge\theta^{c}_{~b}=0~,
\end{eqnarray}
then the curvature $\Omega^{a}_{~b}$ reduces to
\begin{eqnarray}
\label{total-curv} \Omega^{a}_{~b}= R^{a}_{~b}+i\lambda {\tilde
F}^{a}_{~b}~.
\end{eqnarray}

Eq.(\ref{conn00444}) shows that there are two categories of local
gauge transformations: One is the local SU(1,3) group
transformation. The covariant principle naturally leads to the
necessary input of SU(1,3) gauge field; The other is the local
SO(1,3) group transformation. The spin connection represents the
effect of gravitation.

First, let us consider an intrinsic rotation of the daor field
\begin{equation}
\label{daor15} k^{a}\rightarrow k^{\prime a}=l^{\prime a
}_{~b}e^{\prime b}=S^{a}_{~b}k^{b}~,
\end{equation}
where $S^{a}_{~b}$ satisfies Eq.(\ref{ro-trans01}), namely,
$S^{a}_{~b}$ is a faithful representation of SU(1,3) group.
$l^{\prime a }_{~b}$ and $e^{\prime b}$ are defined by $l^{\prime
a }_{~b}=S^{a}_{~c}l^{c}_{d}(S^{-1})^{d}_{~b}$ and $e^{\prime
b}=S^{b}_{~i}e^{i}$ respectively. From reference~\cite{egf80}, it
is known that under the intrinsic SU(1,3) rotation of daor field
the daor connection 1-form $\omega^{a}_{~b}$ transforms as follows
\begin{equation}
\label{daor17} \omega^{\prime
a}_{~~b}=S^{a}_{~c}\omega^{c}_{~d}(S^{-1})^{d}_{~b}+S^{a}_{~c}({\rm
d }S^{-1})^{c}_{~b}~,
\end{equation}
Since $B^{a}_{~b}$ is the connection of SU(1,3) principal bundle,
under the SU(1,3) gauge rotation of the daor field, the
$B^{a}_{~b}$ transforms into
\begin{equation}
\label{daor19+} B^{\prime
a}_{~~b}=S^{a}_{~c}B^{c}_{~d}(S^{-1})^{d}_{~b}+ S^{a}_{~c}({\rm d
}S^{-1})^{c}_{~b}~,
\end{equation}
and $\theta^{a}_{~b}$  satisfies
\begin{equation}
\label{daor1901} \theta^{\prime
a}_{~~b}=S^{a}_{~c}\theta^{c}_{~d}(S^{-1})^{d}_{~b}~.
\end{equation}
Furthermore, it is easy to prove that under this rotation
$F^{a}_{~b}$ and $\Omega^{a}_{~b}$ become
\begin{equation}
\label{daor190302} {F}^{\prime
a}_{~~b}=S^{a}_{~c}F^{c}_{~d}(S^{-1})^{d}_{~b}~,~~{\Omega}^{\prime
a}_{~~b}=S^{a}_{~c}\Omega^{c}_{~d}(S^{-1})^{d}_{~b}~.
\end{equation}

Secondly, consider an orthogonal rotation of the real orthonormal
vierbein
\begin{equation}
\label{daor190304} e^{a}\rightarrow e^{\prime a}=
\Phi^{a}_{~b}e^{b}~,
\end{equation}
where $\Phi^{a}_{~b}$ satisfies
\begin{equation}
\label{daor19030400}
\Phi^{~a}_{c}\eta_{ab}\Phi^{b}_{~d}=\eta_{cd}~.
\end{equation}
Eq.(\ref{daor19030400}) demonstrates that the $\Phi^{a}_{~b}(x)$
is a representation of SO(1,3) group. The doar field transforms as
\begin{equation}
\label{daor1903041} k^{a}\rightarrow k^{\prime a}=l^{\prime a
}_{~c}e^{\prime c}=\Phi^{a}_{~b}k^{b}~,
\end{equation}
where $l^{\prime a }_{~c}=\Phi^{a}_{~b}
l^{b}_{~e}(\Phi^{-1})^{e}_{~c}$. After this transformation, the
new daor connection is
\begin{equation}
\label{daor190305} \omega^{\prime
a}_{~~b}=\Phi^{a}_{~c}\omega^{c}_{~d}(\Phi^{-1})^{d}_{~b}+\Phi^{a}_{~c}({\rm
d }\Phi^{-1})^{c}_{~b}~,
\end{equation}
Similarly, we acquire
\begin{equation}
\label{daor190306} \theta^{\prime
a}_{~~b}=\Phi^{a}_{~c}\theta^{c}_{~d}(\Phi^{-1})^{d}_{~b}+
\Phi^{a}_{~c}({\rm d }\Phi^{-1})^{c}_{~b}~,
\end{equation}
and
\begin{equation}
\label{daor190307} B^{\prime
a}_{~~b}=\Phi^{a}_{~c}B^{c}_{~d}(\Phi^{-1})^{d}_{~b}~.
\end{equation}
The transformation law for the curvatures $R^{a}_{~b}$ and
$\Omega^{a}_{~b}$ under the orthogonal rotation of the real
vierbein is given by
\begin{equation}
\label{daor190308} R^{\prime
a}_{~~b}=\Phi^{a}_{~c}R^{c}_{~d}(\Phi^{-1})^{d}_{~b}~,~~{\Omega}^{\prime
a}_{~~b}=\Phi^{a}_{~c}\Omega^{c}_{~d}(\Phi^{-1})^{d}_{~b}~.
\end{equation}


\section{Daor Field Equations in Empty Space \label{sec:Sec5}}
\renewcommand{\theequation}{5.\arabic{equation}}
\setcounter{equation}{0}

We discuss the daor field formalism for gravity equation in empty
space in this section. Considering the conditions of complexified
field and of reducing the order of the field equations, we propose
the daor field equations in empty space as follows
\begin{eqnarray}
\label{daor7} (\delta^{a}_{~b}{\rm d
}+\omega^{a}_{~b}\wedge)k^{b}=0~,
\\
\label{daor701} \omega_{ab}=\pm \frac{1}{2}~
\epsilon_{abcd}~\omega^{cd}~.
\end{eqnarray}
Where $\epsilon_{abcd}$ is the totally antisymmetric tensor in
4-dimensions~\cite{egf80}. By defining the torsion operator 1-form
\begin{equation}
\label{daor8} \hat{T}\equiv\delta^{a}_{~b}{\rm d
}+\omega^{a}_{~b}\wedge~,
\end{equation}
we can rewrite Eq.(\ref{daor7}) as $\hat{T}k=0$. Multiplying both
sides of Eq.(\ref{daor7}) by the torsion operator $\hat{T}$ yields
\begin{equation}
\label{daor10} \hat{T} \hat{T}k=({\rm d
}\omega^{a}_{~b}+\omega^{a}_{~c}\wedge\omega^{c}_{~b})\wedge
k^{b}=\Omega^{a}_{~b}\wedge k^{b}=0~.
\end{equation}
This equation shows that the operator $\hat{T}$ can be regarded as
the square root of the curvature 2-form.

In Cartan's vierbein method, the Levi-Civita affine connection is
obtained by requiring that the real spin connection $ \theta_{ab}$
satisfies the following conditions
\begin{eqnarray}
\label{daor11} no~torsion&:&~~~~~~{\rm d }
e^{a}+\theta^{a}_{~b}\wedge e^{b}=0~,
\\
\label{daor12} metricity&:&~~~~~~~~~~~~~\theta_{ab}=-\theta_{ba}~.
\end{eqnarray}
$\theta^{a}_{~b\mu}$ is then determined in terms of the vierbeins
and inverse vierbeins and is related to the Levi-Civita affine
connection by
\begin{eqnarray}\label{daor13++}
\theta^{a}_{~b\mu}=-{(e^{-1})}^{\nu}_{~b}e^{a}_{~\nu;\mu}
 =-{(e^{-1})}^{\nu}_{~b}(\partial_{\mu}
e^{a}_{~\nu}-\Gamma^{\lambda}_{~\nu\mu}e^{a}_{~\lambda})~.
\end{eqnarray}
Where $\Gamma^{\lambda}_{~\nu\mu}$ is Levi-Civita affine
connection, which is real and uniquely determined by the
space-time metric
\begin{equation}
\label{Christoffel}
\Gamma^{\lambda}_{~\mu\nu}=\frac{1}{2}g^{\lambda\sigma}(\partial_{\mu}
g_{\sigma\nu}+\partial_{\nu} g_{\mu\sigma}-\partial_{\sigma}
g_{\mu\nu})~.
\end{equation}
Similarly, we extend the metricity and no torsion conditions to
the cases where the total connection is complex. Namely the
complex spin connection $\omega_{ab}$ satisfies
$\omega_{ab}=-\omega_{ba}$ and $T^{a}={\rm d }
k^{a}+\omega^{a}_{~b}\wedge k^{b}=0$. Hence $\omega^{a}_{~b\mu}$
can be expressed in terms of the daor field
\begin{eqnarray}\label{daor13}
\omega^{a}_{~b\mu}=k^{a}_{~\nu}{\bar K}^{\nu}_{~b;\mu} =-{\bar
K}^{\nu}_{~b}k^{a}_{~\nu;\mu} =-{\bar K}^{\nu}_{~b}(\frac{\partial
k^{a}_{~\nu}}{\partial
x^{\mu}}-\Gamma^{\lambda}_{~\nu\mu}k^{a}_{~\lambda})~.
\end{eqnarray}
 It is obvious that Eq.(\ref{daor7}) is generalized torsion-free condition
$T^{a}=0$  and Eq.(\ref{daor701}) implies generalized metricity
condition $\omega_{ab}=-\omega_{ba}$.

Einstein's empty space equation (\ref{daor1}) may be rationally
generalized as follows~\cite{egf80}
\begin{equation}
\label{daor14} \tilde{\Omega}^{a}_{~b}\wedge k^{b}=0~,
\end{equation}
where $\tilde{\Omega}_{ab}$ is the dual of ${\Omega}_{ab}$, which
is defined by
\begin{equation}
\label{dual}
\tilde{\Omega}_{ab}=\frac{1}{2}~\epsilon_{abcd}\Omega^{cd}~.
\end{equation}
The dual of the complex connection is defined by
\begin{equation}
\label{dual++}
\tilde{\omega}_{ab}=\frac{1}{2}~\epsilon_{abcd}\omega^{cd}~.
\end{equation}
From Eq.(\ref{daor1903}) we notice that $\Omega_{ab}$ is
(anti-)self-dual, namely, $\tilde{\Omega}_{ab}=\pm {\Omega}_{ab}$,
if $\omega_{ab}$ is (anti-)self-dual $\omega_{ab}=\pm
\tilde{\omega}_{ab}$. Since Eq.(\ref{daor701}) shows that
$\omega_{ab}$ is (anti-)self-dual, then Eq.(\ref{daor14}) can be
deduced from Eq.(\ref{daor10}).

We have proved that the daor field equations (\ref{daor7}) and
(\ref{daor701}) are equivalent to Eq.(\ref{daor14}). If the
coupling between the SU(1,3) gauge field and the daor field must
be ignored, then Eq.(\ref{daor14}) reduces to Eq.(\ref{daor1}). In
this case spin connection $\omega_{ab}$ is real. This demonstrates
that there are no observable physical phenomena on SU(1,3) gauge
fields. But in more general cases, $\omega_{ab}$ must be complex.
Adding the stress-energy tensor of gauge fields in Einstein's
equation should give the couplings between the daor field and
gauge fields. More results on this problem will be given in the
forthcoming paper~\cite{hua04c}.

It is stressed that only daor field can embody all the symmetries
of space-time. As the complex spin connection unifies gauge fields
and the Cartan's real spin connection, the daor field reflects the
gravitational effect of gauge fields also. Furthermore, the daor
field will be a powerful tool to realize the quantization of
gravity.

\section{Conclusion \label{sec:Sec6}}
\renewcommand{\theequation}{5.\arabic{equation}}
\setcounter{equation}{0}

In this paper, the daor field which represent gravity is
suggested. There are two kinds of symmetry transformations keeping
$ds^{2}$ invariant. Upon these local symmetries we set up the
locally complexified geometry. In the complex connection, the real
spin connection and the SU(1,3) field strength are unified. Hence
we incorporate the SU(1,3) gauge field with the gravitation. In
the last one-order differential equations of the daor field in
empty space are acquired, being proven to be consistent with
Einstein's empty space equation.

{\bf Acknowledgement:} I am grateful to Prof. B. Q. Ma, C. B.
Guan, Y. Q. Li , J. F. Cheng and Y. P. Dai for their useful
discussion.

\end{document}